\documentclass[aps,prb,twocolumn,superscriptaddress]{revtex4}
\usepackage{graphicx}

\begin{document}

\title{Quantum Spin Excitations through the metal-to-insulator crossover in YBa$_2$Cu$_3$O$_{6+y}$}

\author{Shiliang Li}
\email{slli@utk.edu}
\affiliation{
Department of Physics and Astronomy, The University of Tennessee, Knoxville, Tennessee 37996-1200, USA
}
\author{Zahra Yamani}
\affiliation{
Canadian Neutron Beam Centre, National Research Council, Chalk River Laboratories, Chalk River, ON K0J 1J0, Canada
}
\author{Hye Jung Kang}
\affiliation{
NIST Center for Neutron Research, National Institute of Standards and Technology, Gaithersburg, Maryland 20899-8562, USA}
\affiliation{
Department of Materials Science and Engineering, University of Maryland, College Park, Maryland 20742-6393, USA 
}
\author{Kouji Segawa}
\affiliation{
Central Research Institute of Electric Power Industry, Komae, Tokyo 201-8511, Japan
}
\author{Yoichi Ando}
\affiliation{
Institute of Scientific and Industrial Research, Osaka University, Ibaraki, Osaka 567-0047, Japan
}
\author{Xin Yao}
\affiliation{
Department of Physics, Shanghai Jiaotong University, Shanghai 200030, People's Republic of China
}
\author{H. A. Mook}
\affiliation{
Neutron Scattering Sciences Division, Oak Ridge National Laboratory, Oak Ridge, Tennessee 37831-6393, USA
}
\author{Pengcheng Dai}
\email{daip@ornl.gov}
\affiliation{
Department of Physics and Astronomy, The University of Tennessee, Knoxville, Tennessee 37996-1200, USA
}
\affiliation{
Neutron Scattering Sciences Division,
Oak Ridge National Laboratory, Oak Ridge, Tennessee 37831-6393, USA
}

\begin{abstract}
We use inelastic neutron scattering to study the temperature dependence of the spin excitations of a detwinned
superconducting YBa$_2$Cu$_3$O$_{6.45}$ ($T_c=48$ K). In contrast to earlier work on YBa$_2$Cu$_3$O$_{6.5}$ ($T_c=58$ K), where the prominent features in  
the magnetic spectra consist of a sharp collective magnetic excitation termed ``resonance'' and a large ($\hbar\omega\approx 15$ meV) superconducting spin gap, we find 
that the spin excitations in YBa$_2$Cu$_3$O$_{6.45}$ are gapless and have a much broader resonance.
Our detailed mapping of magnetic scattering along the $a^\ast$/$b^\ast$-axis directions at different energies reveals that spin excitations are unisotropic and consistent with the ``hourglass''-like dispersion 
along the $a^\ast$-axis direction
near the resonance, but they are isotropic at lower energies. 
Since a fundamental change in the low-temperature normal state of 
YBa$_2$Cu$_3$O$_{6+y}$
when superconductivity is suppressed 
takes place at $y\sim0.5$ with a metal-to-insulator crossover (MIC), where the ground state transforms from a metallic to an insulating-like phase, our results suggest a clear connection between the large change in spin excitations and the MIC. The resonance therefore is a fundamental feature of metallic ground state superconductors and a consequence of high-$T_c$ superconductivity.

\end{abstract}


\maketitle


\section{INTRODUCTION}

The parent compounds of the high-transition temperature (high-$T_c$) copper oxides are Mott insulators characterized by a very strong antiferromagnetic (AF) exchange in the CuO$_2$ planes and static long-range AF order\cite{orenstein,abanov,scalapino}.  When holes or electrons are doped into the CuO$_2$ planes, the character of the ground state is fundamentally altered from a Mott insulator with static AF order to a superconductor with persistent short-range AF spin correlations (excitations).  If spin excitations are important to the mechanism of superconductivity, they should have universal features for different classes of high-$T_c$ materials and be intimately related to their charge transport properties.  

For YBa$_2$Cu$_3$O$_{6+y}$ (YBCO) with $y \geq 0.5$, the prominent features in spin fluctuations spectra at the superconducting state include a collective magnetic excitation known as the ``resonance'' mode, which is sharp in energy and centered at the AF ordering wavevector $Q$ = (1/2, 1/2), and a superconducting spin gap\cite{birgeneau,tranquada,regnault,fong,dai,hayden,stock,woo}. The magnetic excitation spectra show an ``hourglass''-like dispersion with the resonance at the saddle point\cite{hayden,stock,woo}. The energies of the mode ($E_R$) and the spin gap ($E_{gap}$) track $T_c$'s as $y$ is varied, and the opening of the spin gap below $T_c$ is compensated in part by the spectral weight gain of the resonance\cite{birgeneau,tranquada,fong,dai,hayden,stock}. In the case of optimally doped La$_{2-x}$Sr$_x$CuO$_4$ (LSCO), inelastic neutron scattering measurements reveal that spin excitations also display an hourglass dispersion with a spin gap and incommensurate resonance below $T_c$\cite{birgeneau,tranquada,christensen,vignolle}. Although such behavior is remarkably similar to those of optimally doped YBCO and thus suggests a common microscopic origin\cite{birgeneau,tranquada}, the situation in the  underdoped YBCO and LSCO differs quite dramatically. For deeply underdoped YBa$_2$Cu$_3$O$_{6.353}$ ($T_c$ = 18 K), the spin excitations spectra become gapless and resonance-less; however, they are dominated by commensurate spin excitations and a central diffusive mode associated with a spin-glass phase\cite{sonier,stock2,stock3}. On the other hand, spin excitations in the underdoped La$_{1.875}$Ba$_{0.125}$CuO$_4$ are also gapless and resonance-less but display an hourglass dispersion with incommensurate scattering extending to zero energy\cite{tranquada,tranquada2,xu}. For underdoped superconducting 
LSCO ($0.055\leq x\leq 0.125$), 
spin excitations 
have clear incommensurability down to the lowest doping of 0.055 and 
 behave similarly as those for  
La$_{1.875}$Ba$_{0.125}$CuO$_4$ \cite{yamada}.
Since the resonance appears to be ubiquitous amongst different classes of nearly optimally doped
superconducting copper oxides \cite{fong99,he02,wilson06,zhao07}, it is important to determine its doping evolution
as the mode may be essential to the mechanism of high-$T_c$ superconductivity \cite{demlerzhang,dai99,wilson07}.
The availability of high quality single crystals of YBCO makes this an ideal system for this study.

To understand the microscopic origin of the differences between YBCO and LSCO, we note that although superconductivity in doped copper oxides first appears when the hole-doping level ($x$ in the case of LSCO) exceeds the critical value of 0.055, a fundamental change in the ground state of these materials in the absence of superconductivity actually takes place at much higher doping levels with a metal-to-insulator (MIC) crossover, where the insulating phase under high magnetic fields shows a log (1/$T$) divergence in resistivity\cite{boebinger}. The MIC happens in the underdoped regime for YBCO around $y = 0.5$\cite{ando,sun}, while it occurs near optimal doping for LSCO ( $x\sim$ 0.16)\cite{boebinger}. If spin excitations are important in determining the charge transport properties of doped copper oxides, they might respond to the changes in the electronic correlations across the MIC. 

To test this idea, we judiciously prepared an underdoped YBa$_2$Cu$_3$O$_{6.45}$ [$T_c$ = 48 K, Fig. 1(a)] just below the MIC. Although previous work\cite{regnault,dai} have suggested that the spin gap approaches zero for YBCO with $y$ less than about 0.5, it is unclear that such effect is due to oxygen-doping inhomogeneity in the underdoped regime or associated with a MIC because the difficulty in preparing high quality underdoped single crystals with uniform oxygen concentration and sharp $T_c$.  Compared to previous work on YBa$_2$Cu$_3$O$_{6.5}$ \cite{stock}, which has a 
well-defined sharp resonance and a large ($\hbar\omega\approx 15$ meV) superconducting spin gap, our sample
has slightly less oxygen content and is in the insulating phase of the MIC. 
Our inelastic neutron scattering experiments show that the three key features of the excitations spectra: the spin gap, resonance, and hour-glass dispersion in YBa$_2$Cu$_3$O$_{6.45}$ are all dramatically different from those in
YBa$_2$Cu$_3$O$_{6.5}$ \cite{stock}. Since the MIC for YBCO also occurs near $y=0.5$, our results thus reveal 
a clear connection between the quantum spin excitations and the charge transport properties. 

\section{EXPERIMENT}

For our experiments, we used a solute-rich liquid pulling method to grow a large pure YBCO crystal.
Compared to the earlier well-studied melt-textured bulk YBCO samples, which contain a significant
amount of randomly oriented ``green phase'' (Y$_2$BaCuO$_5$) as an impurity phase \cite{dai}, these new samples 
are ``green phase'' free and 
have no observable impurity phases. The crystal was cut into four pieces with a total mass of 6 grams and the oxygen content was set to $y$ = 0.45 at one atmosphere with 0.5\% oxygen partial pressure at 550 $^{\circ}$C for 5 days. The samples were mechanically de-twinned at 220 $^{\circ}$C and then annealed in sealed tube at 90 $^{\circ}$C for more than three weeks to achieve an ordered oxygen state. 

\begin{figure}
\includegraphics[scale=0.9]{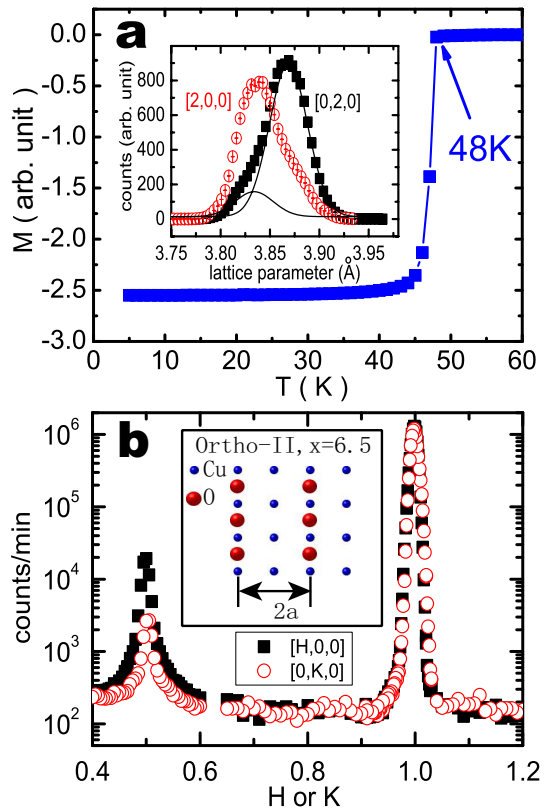}
\caption{(a) The temperature dependence of the magnetic susceptibility showing an onset $T_c$ of 48 K with width of 2 K.  The inset shows (2,0,0) and (0,2,0) Bragg peaks indicating a detwinning ratio of 85\%, as illustrated by the two fitted Gaussian lines for (0,2,0) peak. (b) The ortho-II oxygen ordering (1/2,0,0)/(0,1/2,0) and nuclear (1,0,0)/(0,1,0) peaks. Since the CuO chains form the ortho-II superstructure as shown in the inset, scattering along the $a^\ast$ direction should result in the (1/2,0,0) peak.   
}
\end{figure}

Fig. 1(a) plots the magnetic susceptibility of one of the samples showing an onset $T_c$ of 48 K. 
Based on the empirical $T_c$ vs. oxygen doping plot \cite{liang}, we estimate that the 
sample's oxygen content is $y=0.45$. The inset of Fig. 1(a) shows wavevector scans through (2,0,0) and (0,2,0) Bragg peaks and by fitting these with two Gaussians, we determine the detwinning ratio of 85\%.
This estimated detwinning ratio can also be checked by intensities of copper oxygen chain (CuO) ordering Bragg peaks. Although the oxygen content is lower than 6.5, the copper oxygen chains can still form the ortho-II super structure \cite{andersen}.  Fig. 1(b) shows ortho-II super lattice peaks at (1/2,0,0) and (0,1/2,0) and the intensity ratio between these two peaks confirms the 85\% detwinning ratio.

Since our sample has identical ortho-II CuO chain ordering as that of YBa$_2$Cu$_3$O$_{6.5}$ but with $\sim$10 K lower $T_c$\cite{stock}, a determination of how the sharp resonance and the large superconducting spin-gap in  YBa$_2$Cu$_3$O$_{6.5}$ evolves as the system is tuned to YBa$_2$Cu$_3$O$_{6.45}$ would reveal the evolution of the intrinsic electronic properties across the MIC without complications from changes in the CuO chain anisotropy. For this purpose, we used inelastic neutron scattering experiments to map out the magnetic scattering function, $S(Q,\omega)$, and then obtain the imaginary part of the dynamic susceptibility,$\chi^{\prime\prime}(Q,\omega)$ = $S(Q,\omega)(1-e^{-\hbar\omega/k_BT})$, over a range of energies (0 $\leq \hbar\omega \leq$ 40 meV) on YBa$_2$Cu$_3$O$_{6.45}$ below and above $T_c$.

The low energy spin excitations were carried out on the cold triple-axis spectrometer SPINS at the NIST Center for Neutron Research (NCNR) with a fixed final neutron energy $E_f=5$ meV and a cold Be filter before the analyzer. The collimations are 60$^\prime$-open-Sample-80$^\prime$-open. To study the spin fluctuations at higher energies, we used thermal triple-axis spectrometer C5 at the Canadian Neutron Beam Centre at Chalk River Labs with $E_f$ = 14.7 meV. The collimation for studying resonance and spin dispersion are 30$^\prime$-48$^\prime$-Sample-51$^\prime$-144$^\prime$ and 30$^\prime$-28.6$^\prime$-sample-33.4$^\prime$-144$^\prime$, respectively. We denote positions in momentum space using $Q$ = [$H$, $K$, $L$] in reciprocal lattice units (r.l.u.) in which $Q$ [\AA$^{-1}$] = ($H 2\pi /a$, $K 2\pi/b$, $L 2\pi/c$), where $a = 3.834$, $b = 3.869$, and $c = 11.68$ \AA.

\section{RESULTS}

\begin{figure}
\includegraphics[scale=.7]{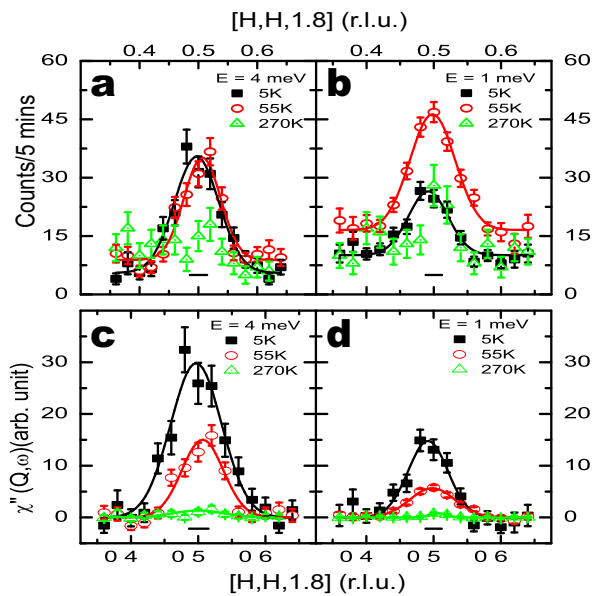}
\caption{$Q$-scans along the $[H,H,1.8]$ direction at (a) $\hbar\omega=4$ meV and (b) 1 meV. While the intensity at 4 meV shows no sign of temperature dependence, the intensity at 1 meV clearly increases when the system changes from 5 K to 55 K. The corresponding $\chi^{\prime\prime}(Q,\omega)$ are shown in (c) and (d), respectively.  The horizontal bars are the instrumental resolutions.
} 
\end{figure}

Since superconducting YBCO in the metallic region of MIC (with $y \geq$ 0.5) have well-defined low-temperature spin-gaps with $E_{gap} \propto T_c$\cite{dai,stock} while spin excitations in the low-$T_c$ YBa$_2$Cu$_3$O$_{6.353}$
 ($T_c=18$ K) are gapless and resonance-less \cite{stock2,stock3}, we first probe the possible elastic (quasi-elastic) AF order and low-energy spin excitations in YBa$_2$Cu$_3$O$_{6.45}$. The purpose of our measurements is to see if YBa$_2$Cu$_3$O$_{6.45}$ has a central diffusive mode similar to those in 
 YBa$_2$Cu$_3$O$_{6.353}$.  

Fig. 2(a) and 2(b) show the $S(Q,\omega)$ in $Q$-scans along  the $[H,H,1.8]$ direction at $\hbar\omega=4$ meV and 1 meV respectively, where $L=1.8$ r.l.u. is the maximum intensity position for the acoustic spin excitations of the bilayer YBCO \cite{regnault,fong,dai,hayden,stock}. It is clear that the excitations at 4 meV show no temperature dependence when the system changes from the low temperature superconducting state ($T = 5$ K) to the high temperature normal state ($T = T_c + 7$ K). However, the $S(Q,\omega)$ peak intensity for $\hbar\omega=1$ meV is larger 
at $T=55$ K than that at 5 K. Fig. 2(c) and 2(d) plot the corresponding $\chi^{\prime\prime}(Q,\omega)$ obtained from data in Figs. 2(a) and 2(b) respectively. The dynamic susceptibility at both energies decrease with the increase of temperatures, which indicates a gapless low-temperature ground state.

\begin{figure}
\includegraphics[scale=0.5]{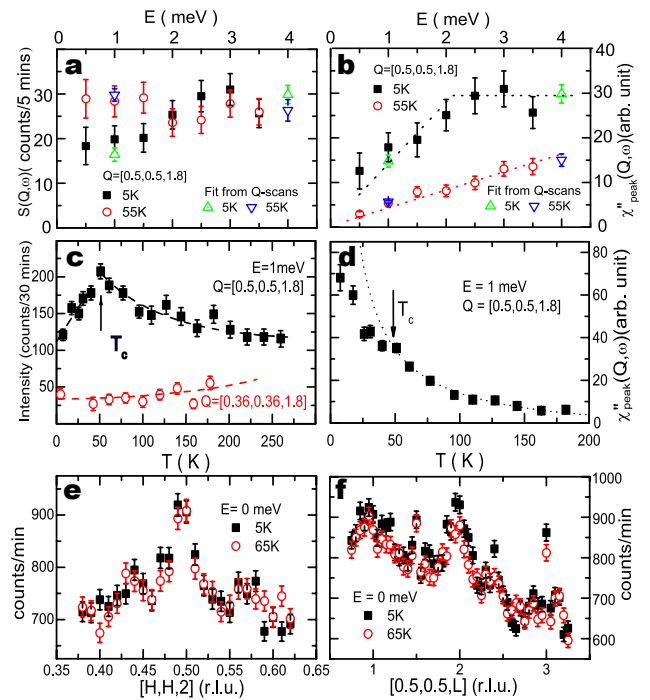}
\caption{(a) Energy scans of the difference between $Q = (0.5,0.5,1.8)$ and (0.36,0.36,1.8) at 5 K and 55 K. The peak intensities obtained by fitting $Q$-scans at 1 meV and 4 meV are also plotted. The corresponding dynamic susceptibilities $\chi^{\prime\prime}(Q,\omega)$ are shown in (b). (c) The temperature dependence of the peak intensity and background at 1 meV. (d) The corresponding $\chi^{\prime\prime}(Q,\omega)$ of (c). The dotted lines are guided to the eye. (e) $Q$-scans along the $[H,H,2]$ direction at $\hbar\omega$ = 0 meV and different temperatures. (f) $L$-scans along the $[0.5,0.5,L]$ direction at $\hbar\omega$ = 0 meV and different temperatures.
}
\end{figure}

To further study the low energy spin fluctuations, we measured the energy dependence of the scattering
at the peak $Q=(0.5,0.5,1.8)$ and background $Q=(0.36,0.36,1.8)$ positions for the acoustic mode \cite{regnault,fong,dai,hayden,stock}. The scattering function $S(Q,\omega)$ at 
different temperatures can then be estimated by 
subtracting the background scattering from the peak position as shown in Fig. 3(a).
Although the statistics of the subtracted data can still be improved, it is clear that the $S(Q,\omega)$ below 2 meV decreases as the system enters into the superconducting state from the normal state.  
Constant-energy scans at $\hbar\omega=1$ and 4 meV below and above $T_c$ shown in Figs. 2(a) and 2(b) 
confirm the result of Fig. 3(a).  Fig. 3(b) shows the $\chi^{\prime\prime}(Q,\omega)$ at 5 and 55 K
obtained using $S(Q,\omega)$ and constant-energy scans in Fig. 3(a).  
While the $\chi^{\prime\prime}(Q,\omega)$ at 55 K increases linearly with $\hbar\omega$, consistent with the energy dependence of the normal state dynamics susceptibility in other metallic ground state YBCO \cite{fong,dai,stock},
it is clear that 
the $\chi^{\prime\prime}(Q,\omega)$ is gapless in the low-temperature superconducting state and shows a 
characteristic energy scale around 2.0 meV.  This energy scale is similar to those found in the deeply underdoped
 YBa$_2$Cu$_3$O$_{6.353}$ \cite{stock2,stock3} and electron-doped superconducting Pr$_{0.88}$LaCe$_{0.12}$CuO$_4$ with $T_c=21$ K \cite{wilsonprl06}.  However, in contrast to  YBa$_2$Cu$_3$O$_{6.353}$ \cite{stock2,stock3}, extensive
 searches in our samples have failed to find any static (or quasielastic) AF order or the central mode. 
Fig. 3(e) gives $Q$-scans along the $[H,H,2]$ direction, which is the position expected for
the central diffusive mode \cite{stock2}. The lack of temperature dependence between 5 K and 65 K indicates that
the scattering is nonmagnetic.  In addition, the $L$-scans along the $[0.5,0.5,L]$ direction shown in Fig. 3(f) do not follow the lattice periodicity as those in YBa$_2$Cu$_3$O$_{6.353}$\cite{stock2}. From previous work on underdoped YBCO with a central mode \cite{stock2,stock3,yamani},
we know that the intensity of the central mode decreases rather rapidly with increasing hole doping and should
certainly be present below 65 K in our YBa$_2$Cu$_3$O$_{6.45}$. A comparison with a recent $\mu$SR results\cite{sonier} suggests that our sample might be close to the magnetic quantum critical point resulting from the spin-glass phase. We also note that the low energy scale around 2.0 meV in our sample is in the superconducting state, different from those in the lower-doping YBCO\cite{stock2,stock3,yamani}.

If the intensity reduction below 2 meV in Fig. 3(a) is indeed related to the bulk superconductivity, 
one should expect that the temperature dependent scattering should respond to superconductivity.
Fig. 3(c) shows the temperature dependence measurements at the peak $Q=(0.5,0.5,1.8)$ and background 
$Q=(0.36,0.36,1.8)$ positions for $\hbar\omega=1$ meV.  The scattering at the peak position initially increases, but then drops substantially below $T_c$ showing a clear kink at $T_c$.  On the other hand, the background scattering 
shows no observable anomaly across $T_c$ as shown in Fig. 3(c). Fig. 3(d) plots the temperature dependence 
of the $\chi^{\prime\prime}(Q,\omega)$ at 1 meV, obtained by subtracting the background from the signal scattering and correcting for the Bose population factor. The $\chi^{\prime\prime}(Q,\omega)$
increases with the decreasing temperature, 
but shows a clear kink at $T_c$.
The fact that the $\hbar\omega$ = 1 meV spin excitations respond to the occurrence of superconductivity effectively rules out the possibility that the low-energy spin excitations arise from the sample oxygen inhomogeneity. In the latter case, one would not expect the low energy spin fluctuations responding to the
bulk superconductivity.

To search for the magnetic resonance in YBa$_2$Cu$_3$O$_{6.45}$, we note that the intensity of the resonance increases below $T_c$ like an order parameter and its energy tracks $T_c$ as the oxygen composition is varied via $E_R$ = 5.8 $k_BT_c$\cite{fong,dai,stock,wilson06}. Since YBa$_2$Cu$_3$O$_{6.45}$ has a $T_c$ = 48 K, we expect the mode to occur at energies around 20 meV. Fig. 4(a) shows energy scans at wavevector $Q = (0.5, 0.5, 5)$ below (5 K) and above (70 K) $T_c$.  Consistent with earlier results on higher-doping YBCO\cite{fong,dai,stock}, the raw data are dominated by phonon scattering at 20 meV and ~30 meV at both temperatures. However, when one takes the temperature difference spectra below and above $T_c$, a broad peak with a full-width-half-maximum (FWHM) of $\sim$ 15 meV emerges at $\hbar\omega \approx$ 19 meV [Fig. 4(c)]. Since intensity of phonons should decrease with decreasing temperature and the Bose population factor does not much affect the magnetic scattering above 10 meV for temperatures from 5 to 70 K, the net intensity gain in Fig. 4(a) must be the result of enhanced dynamic susceptibility below $T_c$.  Although such intensity gain below $T_c$ is a hallmark of the resonance, the observed broad energy peak is quite different from the instrumental resolution-limited resonance for YBCO at higher doping levels\cite{fong,dai,stock}.

\begin{figure}
\includegraphics[scale=0.5]{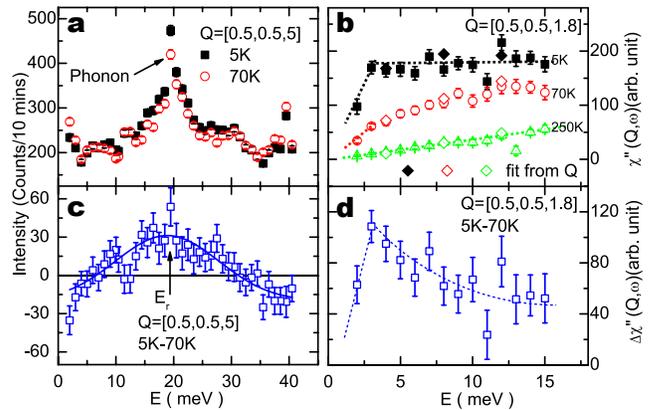}
\caption{(a) Energy-scans at $Q = (0.5,0.5,5)$ taken below (5 K) and above (70 K) $T_c$. The difference in (c) shows a clear peak centered around 20 meV with a FWHM of 15 meV. (b) Temperature dependence of the dynamic susceptibility at the equivalent position $Q = (0.5, 0.5, 1.8)$ showing clear magnetic intensity gain on cooling.  The low-energy dashed lines are from fits in Fig. 3(b). (d) The temperature difference in dynamic susceptibility. 
}
\end{figure}

To see if the intensity gain below $T_c$ is consistent with the bilayer Cu$^{2+}$ acoustic magnetic excitations from YBCO, we carried out energy scans at the equivalent acoustic wavevector $Q = (0.5, 0.5, 1.8)$. The energy dependence of the susceptibilities $\chi^{\prime\prime}(Q,\omega)$ at different temperatures are summarized in Fig. 4(b), where the average values of intensities at $Q=(0.3,0.3,1.8)$ and (0.7,0.7,1.8) have been used as background. Consistent with the cold neutron data in Fig. 3, $\chi^{\prime\prime}(Q,\omega)$ is proportional to $\hbar\omega$ above $T_c$, and increases with decreasing temperature. The difference spectrum $\Delta\chi^{\prime\prime}(Q,\omega)$ between 5 K and 70 K in Fig. 4(d) shows a clear kink around
2 meV in susceptibility consistent with the cold neutron data in Fig. 3(b). Note that the $\chi^{\prime\prime}(Q,\omega)$ is the peak susceptibility at $Q=(0.5,0.5,1.8)$.  To obtain the local susceptibility $\chi^{\prime\prime}(\omega)$ at different energies, one must carry out wavevector integration considering 
 the detailed dispersion of spin excitations.  Although $\Delta\chi^{\prime\prime}(\omega)$ still shows a broad 
peak around 2 meV after considering the $Q$-width below $T_c$, 
the temperature dependence of the magnetic scattering at 2 meV shows no direct correlation with 
$T_c$ and therefore differs from the neutron spin resonance at
$\hbar\omega\approx 19$ meV.

Fig. 5(a)-(c) show wavevector dependence of $\chi^{\prime\prime}(Q,\omega)$ at $\hbar\omega$ = 8, 12, and 24 meV below and above $T_c$. Inspection of Figure reveals that the superconductivity-induced susceptibility gain increases from 12 to 24 meV, and there is substantial magnetic scattering even at 250 K.  
In order to confirm that the observed scattering at 250 K is from acoustic magnetic excitations in YBCO, we carried out $Q$-scans along the $c$-axis direction. The $L$-modulation of the $\hbar\omega$ = 24 meV excitation at 250 K
[inset in Fig. 5(d)] follows the expected acoustic bilayer structure factor, which is proportional to $sin^2(\pi dL)$ ($d$ is the distance between CuO$_2$ planes within a bilayer unit)\cite{fong,dai,stock}. Finally, the temperature dependence of the scattering at $Q = (0.5, 0.5, 5)$ and $\hbar\omega = 24$ meV shows an order parameter-like increase below $T_c$ similar to the temperature dependence of the resonance in the higher-doping YBCO.  These results thus demonstrate that the broad peak centered at $\hbar\omega$ = 19 meV is indeed the magnetic resonance similar to other superconducting cuprates.  

\begin{figure}
\includegraphics[scale=0.5]{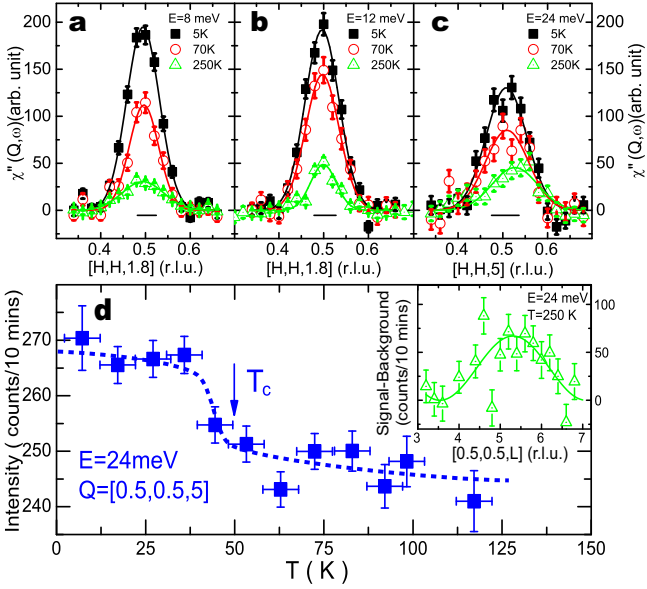}
\caption{ (a)-(c) Wavevector dependence of dynamic susceptibility along the $[H,H]$ direction at different energies and temperatures. There are clear magnetic scattering consistent with acoustic magnetic scattering structure factor even at 250 K, as shown in the inset of (d).  (d) The temperature dependence of the scattering at 24 meV and (0.5,0.5,5) shows an order-parameter-like increase below $T_c$, a hallmark of the resonance.
}
\end{figure}

\begin{figure}
\includegraphics[scale=0.5]{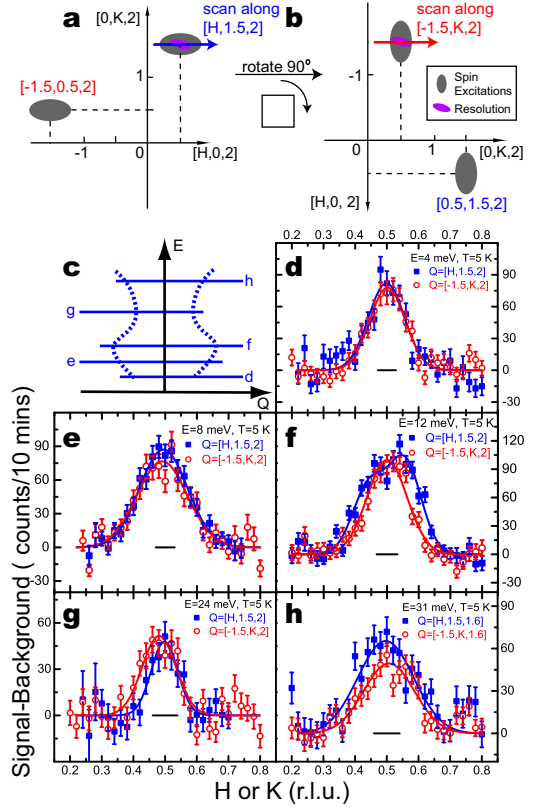}
\caption{(a,b) 
Experimental setup for studying magnetic anisotropy in our YBa$_2$Cu$_3$O$_{6.45}$.
The measurements were carried out in the $[H,K,4K/3]$ and $[H,K,4H/3]$ zones by rotating the sample 90 degrees along the $c^\ast$-axis.  The advantage of such a setup is that the instrumental resolutions are identical in the $a^\ast$ and $b^\ast$ scan directions. (c) Schematic $Q$-scans at different energies.  (d)-(h), $Q$-scans along the $[H,1.5,2]$ and $[-1.5,K,2]$ directions at different energies in the low-temperature superconducting state.  While the scattering profiles are commensurate and isotropic at 4 and 8 meV, clear anisotropic scattering appears at 12 meV. The data show clear flattish top. The $Q$-profiles become narrow and isotropic again near the resonance energy. The instrumental resolutions are shown as the horizontal bars. 
}
\end{figure}

Having shown the presence of the resonance in our YBa$_2$Cu$_3$O$_{6.45}$, it is important to determine its dispersion as the outcome will allow a direct comparison of the magnetic spectra between underdoped YBCO and LSCO. Previous neutron scattering work on YBCO with $y  = 0.5$, and 0.6 have shown that the dispersion of the resonance has an hourglass shape\cite{hayden,stock}, with incommensurate scattering below the resonance being anisotropic, having a magnetic anisotropy with a larger incommensurability along the $a^*$-axis direction than the $b^\ast$-axis direction\cite{mook,hinkov,hinkov2}. Since the low-energy spin fluctuations in our YBa$_2$Cu$_3$O$_{6.45}$ are commensurate, it will be interesting to determine the dispersion of spin excitations along $H$ and $K$ directions near the resonance.  To accomplish this, we co-aligned the samples in either the $[H,K,4/3K]$ or $[H,K,4/3H]$ zone by simply rotating them 90 degrees along the $c^\ast$-axis in the $[H,K,0]$ zone before tilting around the $a^\ast$($b^\ast$)-axis. The unique advantage of such experimental geometries is that one can carry out scans along the $[H,1.5,2]$ or  $[-1.5,K, 2]$ directions with identical instrumental resolution, thus allowing a direct comparison of the possible magnetic anisotropy in this material [Fig. 6(a) and 6(b)].

Fig. 6(d)-6(h) summarize the constant-energy scans along $H$ and $K$ directions for energy transfers of $\hbar\omega = 4$, 8, 12, 24, and 31 meV in the low temperature superconducting state. At $\hbar\omega = 4$ meV, $Q$-scans along the $H$ and $K$ directions show identical behavior and suggest that spin fluctuations are isotropic at this energy [Fig. 6(d)]. On increasing the energy to $\hbar\omega = 8$ meV, the excitations become broader in $Q$ but are still the same along $H$ and $K$ directions [Fig. 6(e)]. Upon increasing the energy further to $\hbar\omega = 12$ meV, the $Q$-scan along the $H$ direction shows a clear flattish top indicative of incommensurate spin excitations while the identical scan along the $K$ direction is commensurate and has a smaller width than the $Q$-scan along the $a^\ast$ direction. We fit the data with two Gaussian peaks with the incommensurability of $\delta = 0.057 \pm 0.003$ r.l.u. being consistent with the universal $\delta$ vs $T_c$ plot\cite{dai}. At energies near and above the resonance (at $\hbar\omega = 24$, and 31 meV, respectively), the scattering profiles become narrow again and the in-plane magnetic anisotropy essentially disappears [Figs. 6(g) and 6(h)]. Our results above 12 meV are consistent with an hourglass dispersion and display a weak anisotropy below the resonance, similar to the previous results at higher doping\cite{mook,hinkov,hinkov2}, but the low energy excitations become isotropic.

To obtain the integrated $\chi^{\prime\prime}(\omega)$, one must carry out two-dimensional wavevector integration
of $\chi^{\prime\prime}(Q,\omega)$ within the $[H,K]$ plane. Since we did not carried out detailed 
$Q$-scans over a wide energy range and it is difficult to separate magnetic scattering from phonons around 20 meV,
we have not attempted to construct a $\chi^{\prime\prime}(\omega)$ versus $\hbar\omega$ plot similar to those for
YBCO \cite{stock,woo} and LSCO \cite{christensen,vignolle}.

\section{DISCUSSIONS and CONCLUSIONS} 

\begin{figure}
\includegraphics[scale=0.5]{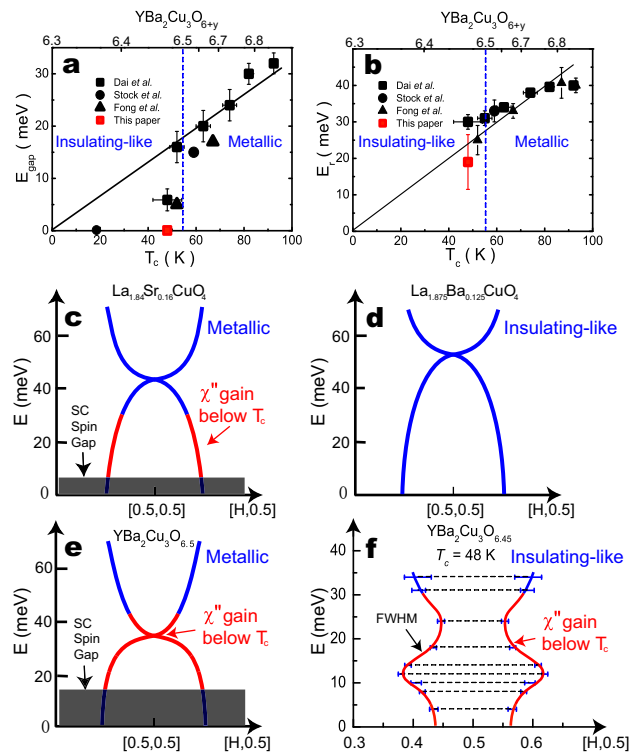}
\caption{(a) Spin gap energies as a function of $T_c$ for YBCO. There is a sudden reduction below $y = 0.5$ and the MIC. (b) The resonance energies and their widths as a function of $T_c$ for YBCO. The energy width of the resonance increases dramatically below $y = 0.5$ and the MIC. (c) The schematic dispersion for optimally doped La$_{1.84}$Sr$_{0.16}$CuO$_4$.  On cooling below $T_c$, a clean spin gap (about 6 meV) opens and the spectral weight shifts from below to above the gap as shown in red region\cite{christensen,vignolle}. (d) The schematic dispersion for static stripe ordered La$_{1.875}$Ba$_{0.125}$CuO$_4$, where incommensurate spin fluctuations extend to zero energy and there is no spectral weight enhancement below $T_c$\cite{tranquada2,xu}. (e) Dispersion of YBCO for $y \geq 0.5$.  The low-energy incommensurate scattering were obscured by the presence of the low-temperature spin gap. The lines within the gap indicate the expected dispersion from a na\"{i}ve stripe model for $y < 0.5$ samples\cite{tranquada}. (f) Observed dispersion for YBa$_2$Cu$_3$O$_{6.45}$ along the $a^\ast$ direction. The part above 12meV is consistent with an hourglass like dispersion, which breaks down at low-energies.
}
\end{figure}

Figure 7 summarizes our present results together with schematics of dispersions from previous work 
on YBCO and LSCO. Figures 7(a) and 7(b)
show the low-temperature spin-gap and resonance energies as a function of $T_c$ for YBCO \cite{fong,dai,stock,stock2}. Figures 7(c-e) 
illustrate the observed spin excitation dispersions for LSCO and YBCO with different ground states. 
Figure 7(f) plots the dispersions of the spin excitations along the $a^\ast$-direction. The disappearing spin gap energy in the sample did not reveal more incommensurate scattering as expected in Fig. 7(e) from a na\"{i}ve stripe picture but instead showed that the low-energy spin excitations are commensurate much different from the dispersion of the lower-doping LSCO. 

The surprising discovery of disappearing spin gap and isotropic spin fluctuations below the hourglass dispersion suggests that the doping evolution of spin excitations in LSCO and YBCO families of cuprates is quite different. The dramatic increase in the magnitude of the spin gap and the width narrowing of the resonance on moving from $y = 0.45$ to 0.5 in YBCO's with ortho-II CuO chain ordering [Fig. 7(a) and 7(b)] suggest that these changes in spin dynamical properties are related with the MIC and its associated quantum phase transition. Since the 
oxygen concentration in YBCO is difficult to determine precisely, the relationship between $T_c$'s and the oxygen contents differ slightly by different groups \cite{ando,sun,liang}. 
The oxygen contents in Figs. 7(a) and 7(b) are calculated based on the empirical results reported by Liang {\it et al.}\cite{liang}. In the original work of Dai {\it et al.}\cite{dai}, spin excitations of an underdoped twinned
melt-textured sample of YBa$_2$Cu$_3$O$_{6.45}$ were found to have a spin gap of 6 meV and a sharp resonance at 30 meV. Although the $T_c$ of this sample was reported to also be $T_c=48$ K, the transition width is about 12 K and
$T_c$ is defined to be the middle point in the DC susceptibility measurement\cite{dai}.  This means that the 
oxygen concentration in the earlier melt-textured YBa$_2$Cu$_3$O$_{6.45}$ is much less homogeneous.  As a consequence, the observed resonance and spin-gap in the melt-textured YBa$_2$Cu$_3$O$_{6.45}$ may arise from portion of the sample with higher $T_c$.  In any case, the spin excitations at 10 meV from the earlier experiments 
also show a tendency of commensurate scattering below the incommensurate peaks at 16 meV \cite{dai}.

Since recent transport measurements have demonstrated the presence of Fermi pockets in the high-field vortex state of YBa$_2$Cu$_3$O$_{6.5}$\cite{leyraud}, we speculate that the sharp resonance and clean spin gap are fundamental properties of metallic ground state copper oxides. Indeed, such speculation is also consistent with the data in LSCO, where one finds a clean spin gap and incommensurate ``resonance'' -- or more precisely, susceptibility gain below
$T_c$ -- near optimal Sr-doping in the metallic phase\cite{christensen,vignolle,tranquada3} but fails to detect any signature of a clean spin gap or resonance in underdoped materials in the insulating-like phase\cite{tranquada2,xu,chang}. 

Our results present new challenges to the two current competing theories explaining the microscopic origin of the spin excitations. In one school of thought, doped holes in the CuO$_2$ are phase separated from the AF insulating background and self-organize into metallic ``stripes'', which necessitates unconventional superconductivity\cite{kivelson2,zaanen}.  Neutron scattering experiments on La$_{1.875}$Ba$_{0.125}$CuO$_4$, where superconductivity is suppressed by the static stripes, have shown that spin excitations form an hourglass dispersion (Fig. 7d) with low-energy incommensurate peaks (measuring spacing between stripes) merge into a saddle point determined by the interactions between stripes\cite{tranquada2,xu}. In this picture, incommensurate spin excitations in the superconducting LSCO and YBCO arise from dynamic stripes\cite{tranquada} and the MIC itself is not expected to much affect the stripe correlations\cite{kivelson2,zaanen}. Our observation of the narrowing commensurate spin excitations below the hourglass resonance in YBa$_2$Cu$_3$O$_{6.45}$ [Fig. 7(f)] is inconsistent with the na\"{i}ve picture of stripe correlations. If stripe correlations were to explain the observed spectra, something more exotic, for example the electronic nematic phase\cite{kivelson}, might be required to become important below the MIC for YBCO\cite{ando,sun}. In fact, the transport measurements in YBCO by Ando {\it et al.}\cite{ando} 
had found an increase in the resistivity anisotropy between the $a$ and $b$ directions
with the decreasing doping attributed to the ``striped'' or electronic nematic phase. Ignoring the quoted 
oxygen concentrations as these may differ from group to group \cite{sun,liang}, we find that 
the resistivity anisotropy becomes much smaller for samples with $T_c\approx 50$ K, consistent with our
observation of isotropic spin excitations at low energies [Fig. 7(f)].
If this scenario is indeed correct, one would expect that the low-energy spin excitations to become anisotropic for YBCO samples with lower $T_c$ and oxygen content.

Alternatively, the resonance and incommensurate spin excitations around it can be explained by quasiparticle scattering across a nested Fermi surface as in conventional metals and superconductors\cite{eschrig,norman}. Based on this approach, a rapid reduction in the magnitude of the spin gap and the near-disappearance of the resonance in the
YBa$_2$Cu$_3$O$_{6.45}$ suggests a sudden change in the Fermi surface topology across the MIC in YBCO\cite{onufrieva}.  Although recent experiments\cite{leyraud} have demonstrated the presence of Fermi pockets in the metallic side of the MIC in the vortex state, it is still unclear what happens to the Fermi surface or if there are any Fermi surfaces at all in the insulating-like regime below the MIC. In any case, a microscopic understanding of the spin excitations\cite{birgeneau,tranquada} and in-plane anisotropy\cite{kivelson2,zaanen} will require new theoretical work that takes into account the differences on emergence of the hourglass dispersions between LSCO and YBCO with increasing doping. Our data provided the missing link in this comparison, and reveal that the resonance and spin gap are fundamental properties of metallic ground state hole-doped superconductors.

Very recently, Rullier-Albenque {\it et al.} \cite{rullier} suggested 
a new microscopic interpretation on the MIC, where 
the differences in out-of-plane lattice disorder between YBCO and LSCO are used to explain why the MIC takes place at different hole-doping levels for these two materials. Since disorder is known to induce in-gap states and low-energy spin excitations \cite{alloul}, one might speculate that the diorder in our YBa$_2$Cu$_3$O$_{6.45}$ is considerably larger than those in the YBa$_2$Cu$_3$O$_{6.5}$ \cite{stock}.  In this picture, the large reduction in the 
spin-gap energy and the broadening of the resonance below the MIC is induced by the sudden increase in the 
lattice disorder in YBa$_2$Cu$_3$O$_{6.45}$.  Although we cannot exclude the possible existence of 
nano-scale defects or intrinsic disorder, the presence of well-established ortho-II order in our sample suggests 
that such defects and disorder do not dominate the Cu-O chain order.  If disorder indeed plays an important role 
in the spin dynamics of YBCO with oxygen content near 6.5, one would expect to observe large changes in spin excitations as the oxygen chain order in the ortho-II YBa$_2$Cu$_3$O$_{6.45}$ is quenched into the ortho-I phase.
Work is currently underway to study the oxygen disorder effects on spin dynamics in YBa$_2$Cu$_3$O$_{6.45}$.

\begin{acknowledgments}
We thank W. J. L. Buyers and Elbio Dagotto for helpful discussions.  This work is supported by the U.S. DOE BES under grant No. DE-FG02-05ER46202.  ORNL is supported by the U.S. DOE under contract No. DE-AC05-00OR22725 through UT/Battelle LLC.  Works at SJTU and CRIEPI are supported by the MOST of China (973 project No. 2006CB601003) and the Grant-in-Aid for Science provided by the Japan Society for the Promotion of Science, respectively.    

\end{acknowledgments}

\end{document}